# USTCSpeech System for VOiCES from a Distance Challenge 2019

*Lanhua You, Bin Gu, Wu Guo*

National Engineering Laboratory for Speech and Language Information Processing,
University of Science and Technology of China, Hefei, China

`{lhyou, bin2801}`@mail.ustc.edu.cn, guowu@ustc.edu.cn

## Abstract

This document describes the speaker verification systems developed in the Speech lab at the University of Science and Technology of China (USTC) for the VOiCES from a Distance Challenge 2019. We develop the system for the Fixed Condition on two public corpus, VoxCeleb and SITW. The frameworks of our systems are based on the mainstream i-vector/PLDA and x-vector/PLDA algorithms.

## 1. Introduction

The USTCSpeech system to Challenge 2019 is based on x-vector&i-vector/PLDA frameworks. We use some open source code as well as our own platform to provide complementarity. Two kinds of acoustic features, PLP and MFCC, are adopted in all the sub-systems. Three i-vector sub-systems are implemented using Kaldi and our own platform. Nineteen x-vectors sub-systems are implemented using Kaldi and Tensorflow platform. The PLDA algorithm is adopted as backend classifier for all the i-vector/x-vectors. The submitted system is a fusion at the score-level of these sub-systems.

The remainder of this document is organized as follows. Section 2 presents the data preparation. In section 3, we describes various sub-systems. After it, section 4 lists results on development set.

## 2. Data Preparation

### 2.1. Individual Datasets

The VOiCES from a Distance Challenge 2019 can be evaluated over a fixed or open training condition. In our work, we only train our model in fixed condition. The datasets used for training included Voxceleb1 & Voxceleb2. The SITW will be used for score normalization.

### 2.2. Data Augmention

The following strategies are used for data augmentation.

- Reverb: The speech utterances are artificially reverberated via convolution with simulated RIRs [2], and we didn't add any additive noise here.
- Music: A single music file (without vocals) is randomly selected from MUSAN corpus, trimmed or repeated as necessary to match duration, and added to the original signal at 5-15dB SNR [3].
- Noise: MUSAN noises are added at one second intervals to the original signal at 0-15dB SNR.

## 3. System Descriptions

We will introduce the sub-systems in this section.

### 3.1. Front-end Modeling

#### 3.1.1. GMM-UBM i-vector

We explored three i-vector [4] systems with different experiment setups.

##### 3.1.1.1 I-vector 1&2

These two sub-systems are implemented using Kaldi toolkit [5].

For the acoustic features, a 25 ms window with 10 ms shifts is applied to compute the 24-dimensional MFCCs and their first and second derivatives. The cepstral filter banks are selected within the range of 20 to 7600 Hz. Short-time cepstral mean subtraction is applied over a 3-second sliding window and energy based VAD is used to remove the non-speech frames then. The features are used to train a 2048 component Gaussian Mixture Model-Universal Background Model (GMM-UBM) with full covariance matrices. After UBM is trained, 600-dimensional total variability matrix is trained using the above-mentioned training set.

We also used PLP features for another sub-system with the same setups. The only difference is that we use 39-dimensional PLP to replace the MFCC.

##### 3.1.1.2 I-vector 3

This sub-system follows the similar setups like i-vector 1, but here we extracted 13-dimentional PLP with their first and second derivatives using HTK toolkit. The i-vector extractor is implemented with our own platform.

#### 3.1.2. Kaldi X-vector 4-6

Three x-vector systems are trained using Kaldi toolkit. The main differences of them are the acoustic feature and network parameters. The training recipe and data are all the same.

For x-vector 4, we extract 30-dimensional MFCCs (including c0) from 25 ms frames every 10 ms using a 30-channel Mel-scale filter bank spanning the frequency range 20 Hz to 7600 Hz. A short-time cepstral mean subtraction is applied over a 3-second sliding window, and an energy based VAD is used to drop the non-speech frames. The parameters of the network are listed in Table 1. The first five hidden layers are constructed with the time-delay architecture and operate at frame-level. Then a statistics pooling layer is employed to compute the mean and standard deviation over all frames for an input segment. The resulted segment-level representation is then fed into two fully connected layers to classify the speakers in the training set. After training, speaker embedding is extracted from the 512-dimensional affine component of the first fully connected layer [6].

Table1. Architecture of Kaldi x-vector

| Layer | Context | Output Size |
|---|---|---|
| 1 | [t-2,t+2] | 512 |
| 2 | {t-2, t, t+2} | 512 |
| 3 | {t-3, t, t+3} | 512 |
| 4 | {t} | 512 |
| 5 | {t} | 1500 |
| pooling | [1,T] | 3000 |
| 6 | - | 512 |
| 7 | - | 512 |

We also trained another 2 sub-systems with similar network architecture. One used the MFCC mentioned above, and the other used 20-dimentional PLP. We find that there is some complementarity between Kaldi x-vectors with different setups.

*3.1.3. Tensorflow X-vector*

The systems introduced in this section are based on the Tensorflow implementation of the x-vector speaker embedding. Some systems are trained using our own Tensorflow code and the others are implemented based on the open source code as described in [7]. Only the training process is implemented in Tensorflow toolkit. The overall steps are same as the original recipe with some modifications which are explained below. The features are extracted with the same setups in 3.1.2. We also extracted several embedding *b* of the sub-systems in this section for system fusion [8].

*3.1.3.1 Tensorflow x-vector using basic architechure =TFXvec 7-9*

The TFXvec 7 is trained using the open source Tensorflow code [7]. Its network architecture is a little different from the Kaldi x-vector 4 where the 1536 output nodes are used for the fifth frame-level layer.

Meanwhile, we trained a sub-system using the same setups with the activation function replaced with PReLU and the context of the third layer replaced with {t-4, t, t+4}. The other sub-system TFXvec 9 is trained where 256 and 2048 output nodes are used for the first four layers and the fifth layer respectively, and the context of the fourth layer is {t-4, t, t+4} in TFXvec 9.

*3.1.3.2 Tensorflow x-vector with Gated CNN layer=TFXvecGcnn 10*

The architecture of TFXvecGcnn 10 is similar to TFXvec 7 except that the first four layers are replaced with the Gated CNN [9] layers.

*3.1.3.3 Tensorflow x-vector with CNN, LRelu and Attention1 = TFXvecCLReluAtt1 11&12*

The system TFXvecCLReluAtt1 11 adopts the attention mechanism. Here we used the same type of attention as what described in [7]. The size of the last hidden layer before pooling is doubled and equally split into two parts. The first part is used for calculating attention weights while the attentive statistics pooling is calculated using the second part.

At the same time, TFXvecCLReluAtt1 11 uses the CNN and LRelu instead of TDNN and Relu.

For TFXvecCLReluAtt1 12, we trained a sub-system using the same type of attention with the 256 output nodes for the first two frame-level layers. The kernel sizes of the CNN layers are 7,5,3,1 and 1 respectively.

*3.1.3.4 Tensorflow x-vector with Attention2 = TFXvecAtt2 13*

In TFXvecAtt2 13, we used the same type of single-head attention mechanism described in [10] for the TFXvec 7.

*3.1.3.5 Tensorflow x-vector with Attention3 = TFXvecAtt3 14*

Unlike TFXvecAtt2, this system directly averages the output of the last frame-level layer to calculate attention weights through softmax function instead of training the additional weighting parameters for calculating attention weights.

*3.1.3.6 Tensorflow x-vector with CNN and He initialization = TFXvecCHeInit 15*

The differences between this system and TFXvec 7 are that CNN is used instead of TDNN and the weight initialization method is He initialization [11] in this system.

*3.1.3.7 Tensorflow x-vector with auto encoder = TFXvecAE 16-20*

In the TFXvecAE systems, the additional task is added in which the network is forced to reconstruct the high-order statistics of input features [12]. The high-order statistics pooling is computed by concatenating the mean, standard deviation, skewness, and kurtosis of input features with a total of 30*4=120 dimensions. The network is trained through multi task learning where the original cross-entropy (CE) loss and mean squared error (MSE) loss are given weights of 0.7 and 0.3 respectively.

We trained other sub-systems TFXvecAE 17-20 with different task weights, shared layers and the input without data augmentation.

*3.1.3.8 Tensorflow x-vector with Gated CNN and Attention =TFXvecGcnnAtt 21&22*

These two sub-systems are trained using our Tensorflow platform. In the TFXvecGcnnAtt 21, the first four layers are constructed with 256-dimensional dilated Gated CNN and the last frame-level layer is modeled by 1500-dimensional CNN. The dilation rates are {1, 2, 4, 1, 1} and the kernel sizes are {5, 3, 3, 1, 1} respectively. The single-head attention statistics pooling [10] is used in the pooling layer. The rest of the structure is the same as that of TFXvec 7.

We also implemented a gate-attention statistics pooling in the second TFXvecGcnnAtt sub-system where the output of the last frame-level layer is modulated by an output gate and the output gate is further used to calculate the attention weights for the attentive statistics pooling layer [13].

**3.2. PLDA backend**

Our PLDA backend is implemented in the Kaldi toolkit [6]. The extracted speaker embeddings (i-vectors or x-vectors) are centered and projected by LDA. The LDA dimension was tuned on the development set. The PLDA model is trained using the longest 200k recordings from the training set and the length-normalization is adopted before the PLDA scoring.

### 3.3. Unsupervised Clustering Score Normalization

We proposed an unsupervised clustering score normalization algorithm for a contrastive system. The motivation of the proposed algorithm is to use the scores from the most competitive impostors for the normalization parameters. Firstly, the K-means clustering algorithm is applied to a large pooling normalization scores. The scores belonging to the clusters with small mean values are discarded and will not be used. Then the expectation maximization (EM) algorithm is applied and the Gaussian mixture models (GMM) are used to fit the distribution of the remaining scores. The parameters of the Gaussian component with the largest mean value are used for normalization.

### 3.4. System Fusion

We submit three systems finally. The scores of most of the sub-systems are fused with equal weights for system 1, which is our primary system. For system 2, the scores of each sub-system are normalized using the above-mentioned unsupervised clustering score normalization. The scoring weight of each sub-system is tuned in development set and the min DCF is used as main criterion. For system 3, the procedure is almost the same with system 2 except that the score normalization is not applied.

## 4. Experiment Results

Results of the final fusion systems for development set are summarized in Table 2.

Table2. Performance summary of the different submitted systems

| System | EER | Min DCF | Act DCF | avgRPrec | Cllr |
| --- | --- | --- | --- | --- | --- |
| System 1 | 0.0246 | 0.2693 | 0.2850 | 0.8741 | 0.3645 |
| System 2 | 0.0248 | 0.2815 | 0.3093 | 0.8722 | 0.3425 |
| System 3 | 0.0228 | 0.2652 | 0.3995 | 0.8755 | 0.2379 |